\documentstyle[epsf]{mn}
\newif\ifAMStwofonts

\def\eps@scaling{.95}
\def\epsscale#1{\gdef\eps@scaling{#1}}

\def\plotone#1{\centering \leavevmode
    \epsfxsize=\eps@scaling\columnwidth \epsfbox{#1}}

\newif\ifAMStwofonts
\def\LCDM{$\Lambda$CDM}
\def\apj{ApJ}
\def\mnras{MNRAS}
\def\apjl{ApJL}
\def\aap{A\&A}

\def\pc{\ {\rm pc}}
\def\Rc{R_{\rm cusp}}

\newcommand{\be}{\begin{equation}}
\newcommand{\ee}{\end{equation}}

\newcommand{\Msun}{{\rm M_\odot}}

\ifoldfss
  \ifCUPmtlplainloaded \else
    \NewTextAlphabet{textbfit} {cmbxti10} {}
    \NewTextAlphabet{textbfss} {cmssbx10} {}
    \NewMathAlphabet{mathbfit} {cmbxti10} {} % for math mode
    \NewMathAlphabet{mathbfss} {cmssbx10} {} %  "   "    "
  \fi
  \ifAMStwofonts
    \ifCUPmtlplainloaded \else
      \NewSymbolFont{upmath} {eurm10}
      \NewSymbolFont{AMSa} {msam10}
      \NewMathSymbol{\upi}     {0}{upmath}{19}
      \NewMathSymbol{\umu}     {0}{upmath}{16}
      \NewMathSymbol{\upartial}{0}{upmath}{40}
      \NewMathSymbol{\leqslant}{3}{AMSa}{36}
      \NewMathSymbol{\geqslant}{3}{AMSa}{3E}

    \fi
  \fi
\fi % End of OFSS

\ifnfssone
  \newmathalphabet{\mathit}
  \addtoversion{normal}{\mathit}{cmr}{m}{it}
  \addtoversion{bold}{\mathit}{cmr}{bx}{it}
  \newmathalphabet{\mathbfit} % math mode version of \textbfit{..}
  \addtoversion{normal}{\mathbfit}{cmr}{bx}{it}
  \addtoversion{bold}{\mathbfit}{cmr}{bx}{it}
  \newmathalphabet{\mathbfss} % math mode version of \textbfss{..}
  \addtoversion{normal}{\mathbfss}{cmss}{bx}{n}
  \addtoversion{bold}{\mathbfss}{cmss}{bx}{n}
  \ifAMStwofonts
    \ifCUPmtlplainloaded \else
      \UseAMStwoboldmath
      \makeatletter
      \new@mathgroup\upmath@group
      \define@mathgroup\mv@normal\upmath@group{eur}{m}{n}
      \define@mathgroup\mv@bold\upmath@group{eur}{b}{n}
      \edef\UPM{\hexnumber\upmath@group}
      \new@mathgroup\amsa@group
      \define@mathgroup\mv@normal\amsa@group{msa}{m}{n}
      \define@mathgroup\mv@bold\amsa@group{msa}{m}{n}
      \edef\AMSa{\hexnumber\amsa@group}
      \makeatother
      \mathchardef\upi="0\UPM19
      \mathchardef\umu="0\UPM16
      \mathchardef\upartial="0\UPM40
      \mathchardef\leqslant="3\AMSa36
      \mathchardef\geqslant="3\AMSa3E
    \fi
  \fi
\fi % End of NFSS release 1

\ifnfsstwo
  \DeclareMathAlphabet{\mathbfit}{OT1}{cmr}{bx}{it}
  \SetMathAlphabet\mathbfit{bold}{OT1}{cmr}{bx}{it}
  \DeclareMathAlphabet{\mathbfss}{OT1}{cmss}{bx}{n}
  \SetMathAlphabet\mathbfss{bold}{OT1}{cmss}{bx}{n}
  \ifAMStwofonts
    \ifCUPmtlplainloaded \else
      \DeclareSymbolFont{UPM}{U}{eur}{m}{n}
      \SetSymbolFont{UPM}{bold}{U}{eur}{b}{n}
      \DeclareSymbolFont{AMSa}{U}{msa}{m}{n}
      \DeclareMathSymbol{\upi}{0}{UPM}{"19}
      \DeclareMathSymbol{\umu}{0}{UPM}{"16}
      \DeclareMathSymbol{\upartial}{0}{UPM}{"40}
      \DeclareMathSymbol{\leqslant}{3}{AMSa}{"36}
      \DeclareMathSymbol{\geqslant}{3}{AMSa}{"3E}
    \fi
  \fi
\fi % End of NFSS release 2

\ifCUPmtlplainloaded \else
  \ifAMStwofonts \else % If no AMS fonts
    \def\upi{\pi}
    \def\umu{\mu}
    \def\upartial{\partial}
  \fi
\fi

\title{The effect of low mass substructures on the Cusp lensing relation}
\author[Andrea V. Macci\`o \& Marco Miranda]
{Andrea V. Macci\`o$^1$ \thanks{E-mail: andrea@physik.unizh.ch} \& Marco Miranda$^1$ \\
$^1$ Institute for Theoretical Physics, University of Z$\ddot u$rich, CH-8057
  Z$\ddot u$rich, Switzerland \\
}
\smallskip 

\pubyear{2005}

\begin{document}

\maketitle

\begin{abstract}
It has been argued that the flux anomalies detected in gravitationally
lensed QSOs are evidence for substructures in the foreground lensing
haloes. In this paper we investigate this issue in greater detail
focusing on the {\it cusp} relation which corresponds to images of a source
located to the cusp of the inner caustic curve.
We use numerical simulations combined with a Monte Carlo approach
to study the effects of the expected power law distribution of substructures
within $\Lambda$CDM haloes on the multiple images.

Generally, the high number of anomalous flux ratios in the cusp configurations
is unlikely explained
by 'simple' perturbers (subhaloes) inside the lensing galaxy, either
modeled by point masses or extended NFW subhaloes. We considered in our
analysis a mass range of $10^5-10^7 \Msun$ for the subhaloes.
We also demonstrate that including the effects of the surrounding
mass distribution, such as other galaxies close to the primary lens,
does not change the results. 
We conclude that triple images of lensed QSOs do not show any direct evidence for dark dwarf
galaxies such as cold dark matter substructure.
\end{abstract}


\section{Introduction}
\label{sec:introduction}
Cold Dark Matter (CDM) simulations predict many more
low mass satellite haloes than are actually observed in the Milky Way (Klypin et
al. 1999, Moore et al. 1999).
It seems that 10-15\% of the mass was left in satellites with perhaps 1-2\% at the projected
separations of 1--2 Einstein radii ($R_e$) where we see most lensed images
(e.g. Zentner \& Bullock 2003, Mao et al. 2004); this is far
larger than the observed fraction of 0.01--0.1\%  in observed satellites (e.g.
Chiba 2002).
Solutions to this mismatch were proposed in three broad classes: satellites are present but
dark if star formation is prevented 
%hide the satellites by preventing
%star formation so they are present but dark 
(Bullock, Weinberg \& Kravtsov 2000), satellites are destroyed due to
self-interacting dark matter (Spergel \& Steinhardt 2000), 
or their formation is prevented by changing the power spectrum to something
similar to warm dark matter with significantly less power on the relevant mass scales 
(e.g. Bode et al. 2001). These hypotheses left the major observational 
challenge of distinguishing dark satellites from non-existent ones.
This became known as the CDM substructure problem.

It has been argued that a possible signature of the presence of dark matter
substructures can be found in strong gravitational lensing of QSOs 
(Mao \& Schneider 1998; Metcalf \& Madau 2001; Chiba 2002; 
Metcalf \& Zhao 2002; Dalal \& Kochanek 2002: Kochanek \& Dalal 2004). 
If a distant image source is close to a cusp 
%AM
(from inside) 
%AM
in a caustic
curve, three of the images will be clustered together and the sum of their
magnifications will be zero (Zakharov (1995), taking the negative parity image to have
negative magnification). This relation holds for a wide class of smooth
analytic lens models (Keeton et al. 2003); on the other hand all known observed
lensed QSOs violate this relation. This has been explained with the presence of
cold dark matter substructures within the lensing galaxy's halo.

However, the discrepancy  found in some systems may be due to
microlensed stars rather than to cold dark matter substructures (Keeton et al.
2003), even if the most peculiar problem is the anomalous
flux ratios in radio lenses.  Radio sources are essentially unaffected by the
ISM of the lens galaxy (see however Koopmans et al. 2003), true absorption appears to
be rare, radio sources generally show little variability 
and most of the flux should come from regions too large to be affected by
microlensing. Therefore dark matter subhaloes appear to be the most likely explanation.

By using low resolution simulation of galaxy formation Bradac et al. (2004)  
claimed that the level of substructures present in simulation
produces violations of the cusp relation comparable to those observed. 
Amara et al. (2004) implanted an idealized model of a galaxy into the center
of a high resolution galactic halo extracted from dissipationless N-Body
simulations to test the effects of substructures on lensed images.
Their findings contrast those of Bradac et al. (2004),
since they found that the substructures produced in a \LCDM~ halo are not abundant
enough to account for the observed cusp caustic violation.
The results of Amara et al. (2004) were also confirmed in a recent work by
Macci\`o et al. (2005). In the later work, in which a fully hydrodynamical simulation of
galaxy formation is used, it is shown that the presence of a dissipative component
greatly enhances the surviving probability of satellites, expecially close to the
center of the galaxy. Nevertheless Macci\`o et al. (2005) also demonstrated that the impact on
lensing of subhaloes in the mass range $10^7 - 10^{10} ~\Msun$ is very small. 
Even with a number of subhaloes about 8 higher of the observed one in this mass
range, the number of multiple lensed QSOs that show a
violation of the cups relation is less than 24\%, in contrast with an observed
one of about 60\%.
This means that if the violation of the cusp relation is due to substructures
inside the primary lens, these must have a mass smaller than $10^7 ~ \Msun$.

The aim of this work is to study the influence of subhaloes with mass 
$10^5 - 10^7 \Msun$ on the cusp relation violation. 
%Such a small mass range for substructures cannot be inspected by the
%resolution actually achievable by N-body/hydro numerical simulations 

The outline of the paper is the following: first we briefly summarize the cusp
relation, then in section \ref{s:iden} we present the lensing numerical
simulations and our modeling for the primary lens, subhaloes and extra-haloes.
Tests for our models and results for three different cusp configurations are presented 
in section \ref{sec:res}. Section \ref{sec:fold} is devoted to a short
discussion on the fold relation. 
A discussion of the results and our conclusions are presented in section \ref{sec:concl}.
Throughout this paper the single large halo that is causing the QSO 
multiple images is referred to as the primary lens. The additional small
scale haloes (inside the host halo) are referred to as subhaloes or substructures. 
Haloes beyond the virial radius of the primary lens are referred to as extragalactic haloes.
We adopt the standard $\Lambda$CDM cosmological
model with the following parameters $\Omega_m=0.3$, 
$\Omega_\Lambda=0.7$, $\sigma_8=0.9$ and $H_o=70$ km s$^{-1}$ Mpc$^{-1}$.

\section{The cusp relation}

There are basically three configurations of four-image systems: fold, cusp,
and cross (Schneider \&  Weiss 1992). In this paper we will mainly 
concentrate on the {\it cusp} 
configuration, that corresponds to a source located close to the cusp of the
inner caustic curve. The behavior of gravitational lens mapping near a cusp
was first studied by Blandford \& Narayan (1986), Schneider \& Weiss (1992),
Mao (1992) and Zakharov (1995), 
who investigated the magnification properties of the cusp images
and concluded that the sum of the signed magnification factors of the three
merging images approaches zero as the source moves towards the cusp. In other
words (e.g. Zakharov 1995) :
\begin{equation}
R_{cusp} = {{ \mu_A + \mu_B + \mu_C} \over { \vert \mu_A \vert + \vert 
\mu_B \vert + \vert\mu_C \vert}} \rightarrow 0, ~~for ~~~~\mu_{tot} ~\rightarrow \infty
\label{eq:cusp1}
\end{equation}
where $\mu_{tot}$ is the unsigned sum of magnifications of all four images,
and A,B \& C are the triplet of images forming the smallest opening angle (see
figure \ref{fig:conf1}). 
By opening angle, we mean the angle measured from the galaxy center and being
spanned by two images of equal parity. The third image lies inside such an angle.
This relation is an asymptotic relation and holds when the source approaches
the cusp from inside the inner caustic ``astroid''.
This can be shown by expanding the lensing map to third order in the angular
separation from a cusp (Schneider \&  Weiss 1992). 
%AM
Small scale structure on scales smaller than the image separation
%on approximately the scale of the image separations 
%AM
will cause $R_{\rm cusp}$ to differ from zero
fairly independently of the form of the rest of the lens. Indeed, a substructure 
is more likely to reduce the absolute magnification for negative
magnification images (Metcalf \& Madau 2001, Schechter \&  Wambsganss 2002,
Keeton et al. 2003) and to increase it for positive parity images.

\section{Lensing Simulations}
\label{s:iden}

We use the {\it lensmodel} package (Keeton 2001)\footnote{The software is
public available via web site: http://cfa-www.harvard.edu/castles}
modeling the main lens galaxy as a singular isothermal ellipsoid (SIE) and the 
substructures as NFW (Navarro, Frenk \& White 1997) haloes.
First, using the {\it gravlens} task, we find three lens configurations for which 
the cusp relation is roughly satisfied 
(figures \ref{fig:conf1}, \ref{fig:conf2} and \ref{fig:conf3}). 

As second step a variable number of substructures is added to the main lens (see
section \ref{subh} for details on their number density and physical properties).
For this new lensing system (main lens plus subhaloes) we
compute again positions and fluxes of the images (subhaloes mainly tend to
modify fluxes more than positions, see Kochanek (2004) and section \ref{sec:test}), 
obtaining a new value for the cusp relation $\Rc$.
This procedure is repeated more than 20.000 times for each of three studied
positions of the source (figures \ref{fig:conf1}, \ref{fig:conf2}
and \ref{fig:conf3}): this allows us to compute the probability distribution of the $\Rc$
value in presence of subhaloes (i.e. figure \ref{fig:Multisub1}).
%AM
%In the following we will better describe our parameterization of the main lens
%and subhaloes, with particular attention for the expected number density (from
%N-body simulations) of the latter ones in the mass range considered in this work.

\subsection{Primary Lens}

The observed discrepancy in the flux ratios, compared with the expected
universal relation from a cusp or fold singularity, suggests that it is an
intrinsic difficulty for smooth lens models, not associated with a
particular parameterization. For the scope of this paper it is sufficient to
choose just a single smooth lens model for the primary lens.
Therefore, we select, as a smooth lens model, a singular isothermal ellipsoid
(SIE) (Kormann, Schneider, \& Bartelmann 1994) to take advantage of its
simplicity. 
This model has been widely used in lens modeling and successfully
reproduces many lens systems (e.g. Keeton et al. 1998, Chiba
2002, Treu \& Koopmans 2004). 
An isothermal profile for the total mass distribution of elliptical
galaxies is well supported by the detailed dynamical studies of local ellipticals
(Gerhard et al. 2001), individual lens modeling, and statistics
(e.g. Maoz \& Rix 1993; Kochanek 1995; Grogin \& Narayan 1996).

The ellipsoidal primary lens has a mass equal to $5 \times 10^{11} \Msun$, it is
oriented with the major axis along the y axis in the lens plane and has an
ellipticity of 0.33.
The redshifts of the lens and he source are fixed to $z_l=0.3$ and $z_s=1.71$
respectively, agreement with the typical observed ones (in this case we use
PG1115+080 data, see Tonry 1998)

\subsection{Subhaloes}
\label{subh}

Since it has been shown that the number density of subhaloes with mass $M>10^7 \Msun$ is 
not sufficient to explain the observed number of violation in the cusp relation
(Amara et al. 2004, Macci\`o et al. 2005), the aim of this work is to investigate the
impact of substructures below this mass threshold that is fixed by the
current resolution limits of numerical simulations.

We would like to emphasize that we are using only one lensing plane, this means 
that we will consider only effects due to substructures
being at the same redshift of the main lens (see Chen, Kravtsov \& Keeton
2003, Metcalf 2005 for an estimation of the effects of haloes along the line of sight).
In order to evaluate the number density substructures in the mass range
$10^5-10^7 \Msun$ we have made some extrapolations based on results from high
resolution N-body simulations. 
The mass function of subhaloes inside the virial radius of an halo is close to
a power law (Diemand et al. 2004, DMS04 hereafter, Gao et al.
2004, Reed et al. 2005):
\be
N(>m) \propto m^{-{\beta}},
\label{eq:Nm}
\ee
with a slope $\beta \approx 1$, so that we expect to have a factor $\approx 100$ more
subhaloes inside the viral radius if we move our mass threshold from $10^7
\Msun $ to $10^5 \Msun$. 

\begin{figure}
\plotone{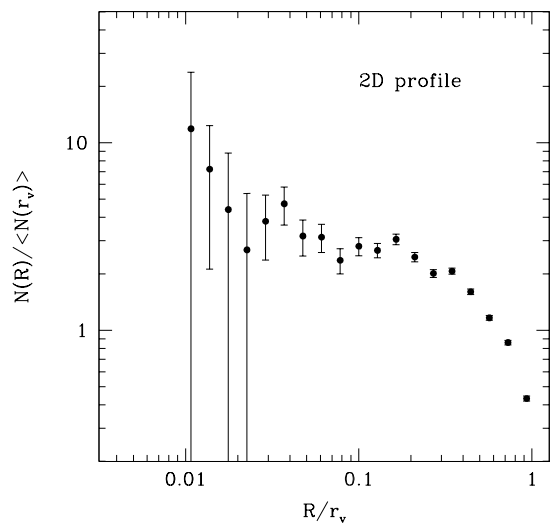}
\caption{Two dimensional radial number density of subhaloes in units of the
  average number density inside the virial radius. 
  Result obtained averaging the 4 high resolution galaxies presented in DMS04,
  using three different projections for each galaxy.}
\label{fig:Nprof}
\end{figure}

As said in the previous section such small haloes will affect the $\Rc$
relation only if their distance from the images is of the same order or
%AM
smaller than the distance between the images themselves. 
%AM
Therefore we need an estimation of the
number of haloes inside a small area surrounding the images. This number will
also depend on the distance of our area from the center, due to the fact that
the number density of haloes increases approaching the center of the main
halo (primary lens) as clearly shown in fig \ref{fig:Nprof}, which is based on
numerical simulations of 4 galaxy size haloes (DMS04).

Consequently the number of subhaloes with a mass greater than $m$ inside an area $A$ at a
distance $R$ from the center of the galaxy is:
\be
N_A(>m,R) = {{ \langle N_{r_v}(>m_0) \rangle  { {m_0} \over {m}}  N(R) A} 
  \over {\pi r_v^2}} ,
\label{eq:Nr}
\ee
where $\langle N_{r_v}(>m_0) \rangle$ is the average number density of subhaloes
with $m>m_0$ (being $m_0$ an arbitrary mass value) inside the virial radius
$r_{v}$ and $N(R)$ is the radial 2D number density of satellites at a
projected distance $R$ from the center in units of $\langle N_{r_{v}}(>m_0) \rangle$ (see
figure \ref{fig:Nprof}). 
These last two quantities can be obtained directly from N-body simulations. 
In the following we will use results from DMS04 (table 1 of their paper, simulations G0-G3).
Macci\`o et al. (2005) have shown that the presence of baryons inside subhaloes enhance
the probability to find haloes close to the center of the galaxy with respect to
results from dissipationless simulations. This is true for satellites with
$m > 5 \times 10^7 \Msun$, which are massive enough to retain baryons inside their
potential well and then form stars. Since we do not expect such effect for the mass
scales involved in this work ($\approx 10^5-10^6 \Msun$) we can use pure
N-body simulation as starting point for our analysis.

Since the typical separation between images in lensed QSOs is roughly a few arcsec,
we fix $R \approx 1$ arcsec and $A=6$ arcsec$^2$ ($A_6$ hereafter).
%AM
We remind that at a redshift of $z_l=0.3$ one arcsec corresponds to 4.55 kpc for the
cosmological model adopted in this paper.   
%AM
Using eq. \ref{eq:Nr} and
adopting a mass threshold for substructures of $m = 5 \times 10^5 \Msun$  
the number of subhaloes inside $A$ ranges from 4 to 12 (depending on the uncertainties on
$N(R)$, see figure \ref{fig:Nprof}).
For two $10^7 \Msun$ haloes the surface mass density within the
selected area $A$ is $0.69 h^{-2} \Msun \pc^{-2}$, this means a fraction $\approx
10^{-3}$ of the total dark matter surface density in substructures, 
in good agreement with the results of Mao et al 2004.

For each lensing configuration analyzed in this work we added 
a random number of substructures between $4-12$ to the primary
lens with a random mass generated
according to eq. \ref{eq:Nm} in the range $5\times 10^5 - 10^7 \Msun$.
These subhaloes are then placed following the 2D density profile inside the
area $A_6$ that encloses the three images (cfr the (blue) square in figure \ref{fig:conf1}).
We have modeled our subhaloes with an NFW (Navarro, Frenk \& White 1997) density
profile; for the $\approx 10^6 \Msun$ subhaloes relevant for lensing substructure studied in
this work, the NFW profile inferred from N-body simulation is the most natural
choice, because on these mass scales the effect of baryons (that are able to
modify the slope of the density profile for greater masses (Macci\`o et al. 2005))
is very tiny because the potential well of these haloes is not deep enough to
retain them expecially in presence of a ionizing background.
We have adopted different concentration parameters (see sec. \ref{sec:test}) to mimic 
the scatter present in the mass-concentration relation (Bullock et al. 2001).

%%%%%%%%%%%%%%%%%%%%%%%%%%%%%%%%%%%%%%%%%%%%%%%%%%%%%%%%%%%%%%%%%%%%%%
%%%%%%%%%%%%%%%%%%%%%%%%%%%%%%%%%%%%%%%%%%%%%%%%%%%%%%%%%%%%%%%%%%%%%%

\subsection{Extrahaloes}
\label{Exth}

Lensing galaxies are not isolated object, since they usually belong to group of
galaxies (Keeton et al. 2000). 
Moreover each galaxy has its own satellites galaxies with masses in
the range $10^9-10^{11} \Msun$. 
Consequently we have modeled the presence of these {\it extrahaloes} in the same way of
the {\it innerhaloes}. In this work we call substructures or innerhaloes, haloes with 
mass $<10^7 \Msun$ that are close to the image position; we reserve the term
extrahaloes for haloes with $M>10^9 \Msun$. We do not consider extrahaloes that for 
projection effects can lie inside the primary lens close to the images
positions (i.e. Oguri 2005). 

We considered three different categories of extrahaloes: i) haloes with mass
$10^9<M< 10^{10} \Msun$ and with a projected distance $r$ between $60$ and $200$
kpc, these represent the satellites galaxy of the primary lens
(the expected number for these haloes can be estimated again form N-body
simulation and it is roughly 6-8); ii) haloes with mass $10^{11}<M< 10^{12} \Msun$ and 
distance $300<r<700$ kpc in order to mimic the presence of companion galaxies;
iii) haloes with $10^{12}<M< 5 \times 10^{13} \Msun$ and $700<r<1200$ kpc to 
take into account the possible presence of a nearby cluster of galaxies.

The number of extrahaloes has been fixed between 2-8 and 2-4 for the ii)
and iii) case respectively, as suggested by observations/simulations (Metcalf 2005,
Amara 2004). 
While the extrahaloes in case i) are placed in a circularly symmetric way around the 
center of the galaxy, the position of groups and clusters of galaxies must be 
modeled in an asymmetric way. 
Therefore we placed them only in the quadrant with positive coordinates in the lens
plane (being the lens in $[0:0]$).
We used SIE as lens model for extrahaloes to take into
account the presence of baryons their-inside and we have generated 20.000
different configurations.

\section{Results}
\label{sec:res}

In this section we present results of our Monte Carlo simulations. We have
analyzed 3 cusp configurations: Config1 (figure \ref{fig:conf1}), Config2 (figure
\ref{fig:conf2}) and Config3 (figure\ref{fig:conf3}). They mainly differ for the
value of $R_{cusp}$ in the unperturbed case that grows from 0.01 for Config1 to
0.243 for Config3, due to a different position of the source inside the inner
caustic curve.

\subsection{Testing our model}
\label{sec:test} 

Before proceeding further in our analysis we present some tests on the
parameter adopted in our lensing simulation. The presence of substructures
acts both on positions and fluxes of the images. 
When a substructure is added the {\it gravlens} code adjusts the 
positions and fluxes of the images by minimizing the $\chi^2$ between the old (unperturbed)
and new (perturbed) positions/fluxes.
The value of $\chi^2$ depends on the error that we assign to the
unperturbed positions and fluxes (usually these are the observational errors); 
this means that if the error on
positions is smaller than the error on fluxes the code will change the
latter ones more than moving the images to obtain a lower value for $\chi^2$.

\begin{figure}
\centering
\plotone{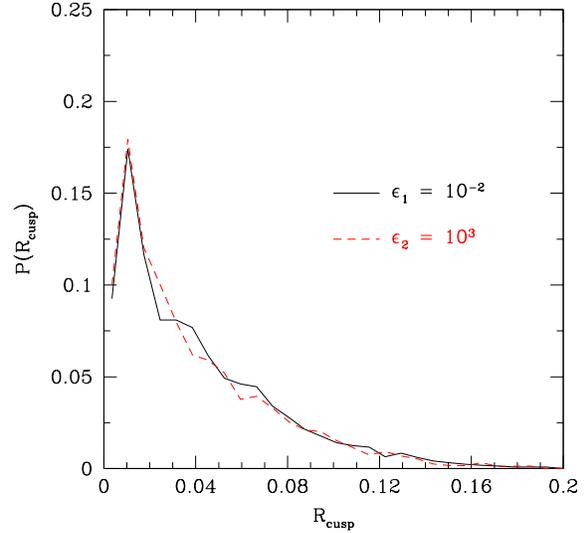}
\caption{Probability distribution of $\Rc$ values for 2 substructures inside
  $A_3$ for different position errors.}
\label{fig:MultisubErr}
\end{figure}

Therefore in principle the value of the perturbed $\Rc$ is influenced by the 
error assigned to the position of the images. In figure \ref{fig:MultisubErr} 
we clearly show that this effect is very small even for a big variation of the
image position errors. This plot illustrates the probability distribution 
for $\Rc$ (obtained using 20.000 realizations) when 4 substructures are
added to the primary lens : 
the solid line is for an error on positions of
$\epsilon_1=10^{-2}$ arcsec ($\approx 5\%$) the dashed one for
$\epsilon_2=10^{3}$ arcsec (i.e. the code has complete freedom in moving the
images).
The two distributions of the $\Rc$ values are very similar, and this also confirms
findings of other authors: the influence of substructures on the
image positions is less strong than the one on fluxes (Kochanek 2004).
We will adopt $10^{-2}$ arcsec for the error on positions for all our simulations.

%**********************************************************************
%**********************************************************************

\begin{figure}
\centering
\plotone{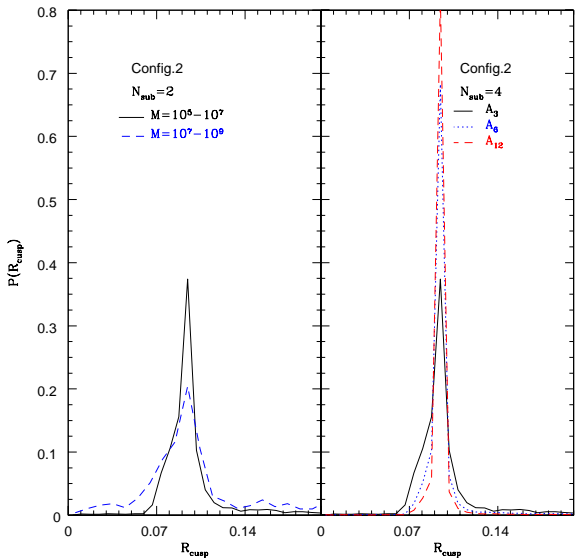}
\caption{Probability distribution of $\Rc$ with 4 substructures: left panel
  influence of the substructure mass. Right panel effects of the projected number
  density.}
\label{fig:MassArea}
\end{figure}

A correct determination of the subhaloes properties
(see section \ref{subh}) is a key ingredient in computing the $\Rc$ value:
figure \ref{fig:MassArea} shows the influence of the number density and mass range of
substructures on the $\Rc$ relation. 
Both changing the area in which subhaloes
are distributed 
%AM
(keeping fixed their number: left panel) 
%AM
or their mass range (right panel) leads to
complete different results. 

%**********************************************************************
%*************************************************************************

\begin{figure}
\centering
\plotone{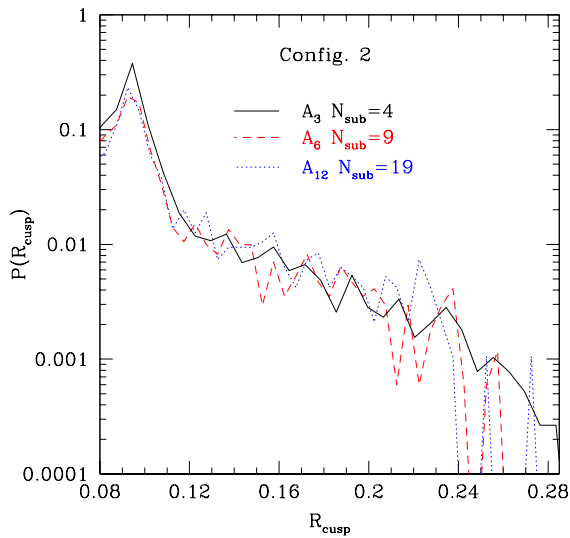}
\caption{Probability distribution of $\Rc$ for substructures distributed on
  different areas.}
\label{fig:RcDiffArea}
\end{figure}

As second  step we have tested if the subhaloes that live outside the small area ($A$)
surrounding the images can substantially modify the $\Rc$ relation.
For this purpose we have defined three different areas for substructures for Config2:
$A_3=[-1.5:1.5]\times[1:2]$, $A_6=[-1.5:1.5]\times[0:2]$ and 
$A_{12}=[-1.75:1.75]\times[-1.25:2.25]$. The last one ($A_{12}$) is big enough to cover 
all the image positions. Changing the size of the area and his position in respect to
the center of the lens, the number of substructures changes according to eq:
\ref{eq:Nr}. The number of subhaloes inside the three areas is 4, 9 and 19
respectively.
Figure \ref{fig:RcDiffArea} shows our results. The $\Rc$ probability
distribution is weakly affected by the size of the area. This indicates that
are the subhaloes close to the image positions the ones responsible for the 
$\Rc$ relation modification.

\begin{figure}
\centering
\plotone{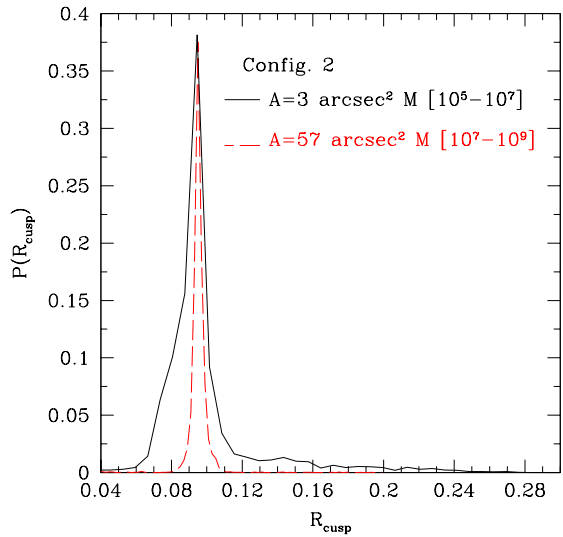}
\caption{Effects on the $\Rc$ relation of substructures inside a small area
($A_3$) surrounding the images (solid line) and in a larger area $[-5:5]\times [0:6]$
  (57 arcsec) outside $A_3$ (dashed line). In the first case the subs are
  in the mass range $[5 \times 10^5:10^7 \Msun]$ in the second one $[10^7:10^9
  \Msun]$.}
\label{fig:outside}
\end{figure}

Figure \ref{fig:outside} shows results for Config2 (figure \ref{fig:conf2}) 
of the $\Rc$ probability distribution for 3 subhaloes with 
mass $5 \times 10^5 < m < 10^7 \Msun$ inside $A_3$
(solid line) and for 8 subhaloes with $10^7 < m < 10^9 \Msun$ inside an area of 57
arcsec$^2$ ($[-5:5]\times [0:6]$ with the exclusion of $A_3$). 
For this second population of substructures both the number density and the
masses are over estimated by a large fraction.
Nevertheless its effect on the cusp relation is very small and the $\Rc$
probability distribution is close to a delta function centered on the unperturbed
value.

%**********************************************************************
%\begin{figure}
%\centering
%\plotone{fig20.ps}
%\caption{Distribution of 12 substructures in an area of $\approx$27
%  Kpc$^2$. In this case we have R=0.238}
%\label{fig:DiffArea}
%\end{figure}
%*************************************************************************
%**********************************************************************
%*************************************************************************
%*************************************************************************

\begin{figure}
\centering
\plotone{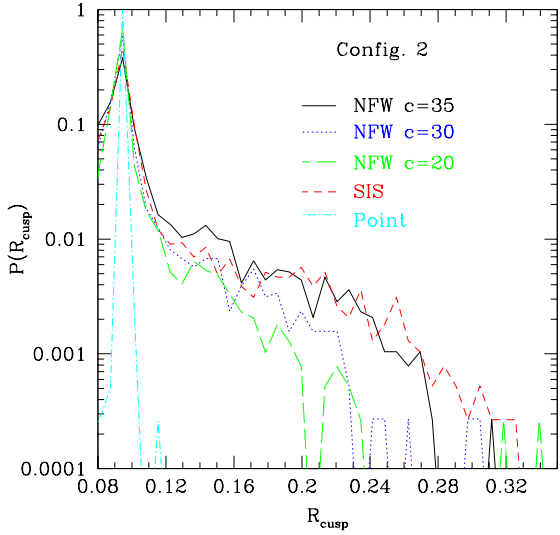}
\caption{$\Rc$ distribution probability for different subhaloes mass profiles: NFW
  profile with different concentration values, SIS profile and point-like approximation.
}
\label{fig:NFWc}
\end{figure}

Modeling our subhaloes as NFW structures we have one more free parameter: the
concentration.
It is well know that the concentration correlates with the mass of the halo,
even if a consistent scatter is present in this relation.
Extrapolating results from Bullock et al 2001, the mean concentration in our
mass range is around 33. In order to test the influence of the concentration
of our subhaloes in modifying the $\Rc$ relation we repeated our analysis
keeping fix mass and position of the subhaloes and varying their
concentrations.
Results are shown in figure \ref{fig:NFWc}. As expected denser haloes have a
stronger impact on the $\Rc$ relation. We have also considered two other mass
profiles for the subhaloes: SIS (singular isothermal sphere) and point like
approximation. The first one, less favored by simulations, can be seen as an
upper limit for the NFW profile, due to the fact that its inner slope for the
density profile is proportional to $r^{-2}$. For this kind of subhalo model
the $\Rc$ probability is just slightly above the ones for NFW haloes with c=35,
so we do not expect a big change in our results using SIS instead of NFW subhaloes.
On the other hand the point mass approximation 
leads to an underestimate of the effect of subhaloes on $\Rc$  
since for a fixed mass a point-like object has a smaller Einstein radius
than an NFW halo (Keeton 2003).
In the following we have adopted a concentration parameter of 35 for our
subhaloes in order to try to maximize their effect on the $\Rc$ relation.

%%%%%%%%%%%%%%%%%%%%%%%%%%%%%%%%%%%%%%%%%%%%%%%%%%%%%%%%%%%%%%%%%%%%%%%%%%%%%%%%%%
%%%%%%%%%%%%%%%%%%%%%%%%%%%%%%%%%%%%%%%%%%%%%%%%%%%%%%%%%%%%%%%%%%%%%%%%%%%%%%%%%%

\subsection{Configuration 1}

\begin{figure}
\centering
\plotone{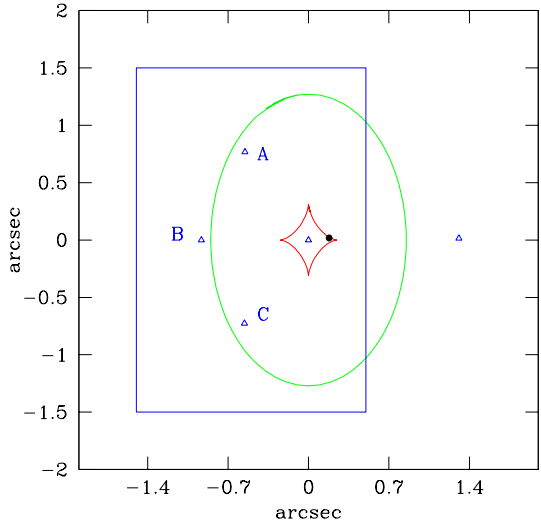}
\caption{Unperturbed lens configuration: Config1 (R=0.01)}
\label{fig:conf1}
\end{figure}

In the first configuration analyzed (Config1 hereafter) the source is close to the right cusp
of the inner caustic curve (see fig \ref{fig:conf1}) and the unperturbed $\Rc$
value is 0.01. 
%AM
The critical and caustic curves refer to the unperturbed case but they
are not different from perturbed ones at the level of resolution.
%AM

\begin{figure}
\centering
\plotone{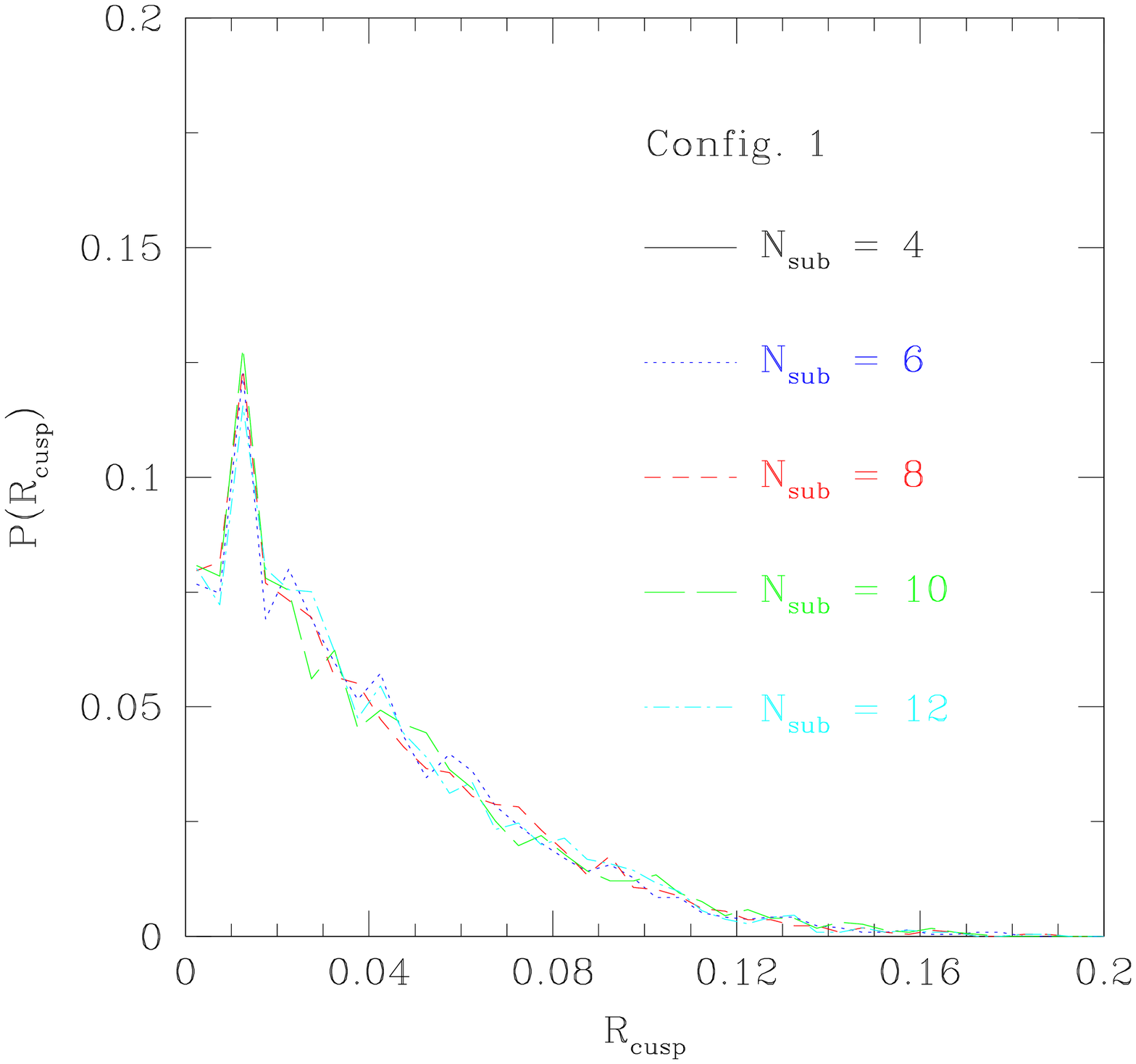}
\caption{Probability distribution of R-variation for a different number of substructures
  inside $A_6$.}
\label{fig:Multisub1}
\end{figure}

\begin{figure}
\centering
\plotone{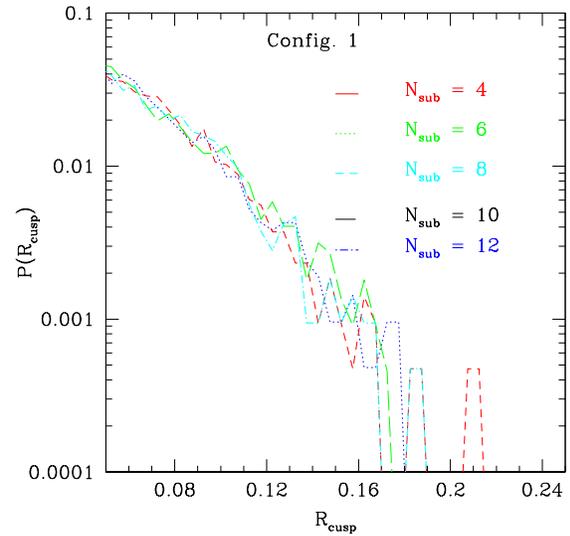}
\caption{Same of figure \ref{fig:Multisub1} with logarithmic scale for the y axis.}
\label{fig:Multi1big}
\end{figure}

For this configuration we have generated 20.000 different lensing systems
that include substructures according to eq: \ref{eq:Nr}. In figure 
\ref{fig:Multisub1} is shown the probability distribution for $\Rc$ for 
different numbers of substructures. The maximum of the probability is obtained
for the unperturbed value (0.01) and the tail of the distribution extends
to $\Rc=0.12$ but for a very low number of configurations (less
than 1.0\%). Figure \ref{fig:Multi1big} shows in a logarithmic plot the tail
of the distribution presented in figure \ref{fig:Multisub1}: it is possible
to note that an increase of the total number of substructures (from 4 to 6) 
does not substantially change the value of $\Rc$.

From figure \ref{fig:Multisub1} one see that in the 10\% of the
configurations the final value of $\Rc$ is even less than the unperturbed value
and it is closer to the theoretical expectation of $\Rc=0$ (see next 
section on the second configuration for more details).

%**********************************************************************
%**********************************************************************
%**********************************************************************
%**********************************************************************

\subsection{Configuration 2}

\begin{figure}
\centering
\plotone{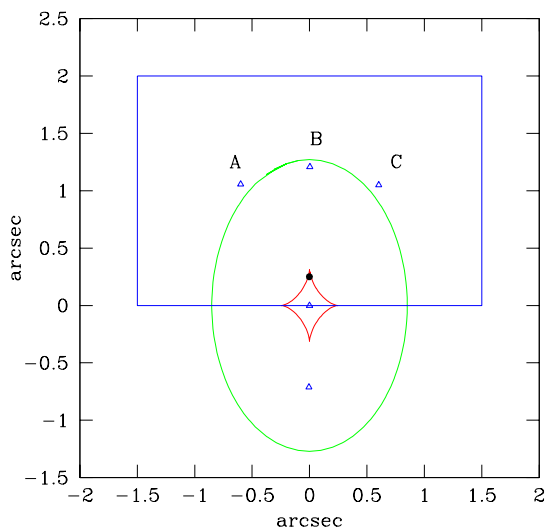}
\caption{Second unperturbed lens configuration: Config2 (R=0.09)}
\label{fig:conf2}
\end{figure}

In this configuration (Config2 hereafter) the source is close to the upper cusp
of the inner caustic curve (see fig \ref{fig:conf2}) with an unperturbed value
of $\Rc=0.09$.

\begin{figure}
\centering
\plotone{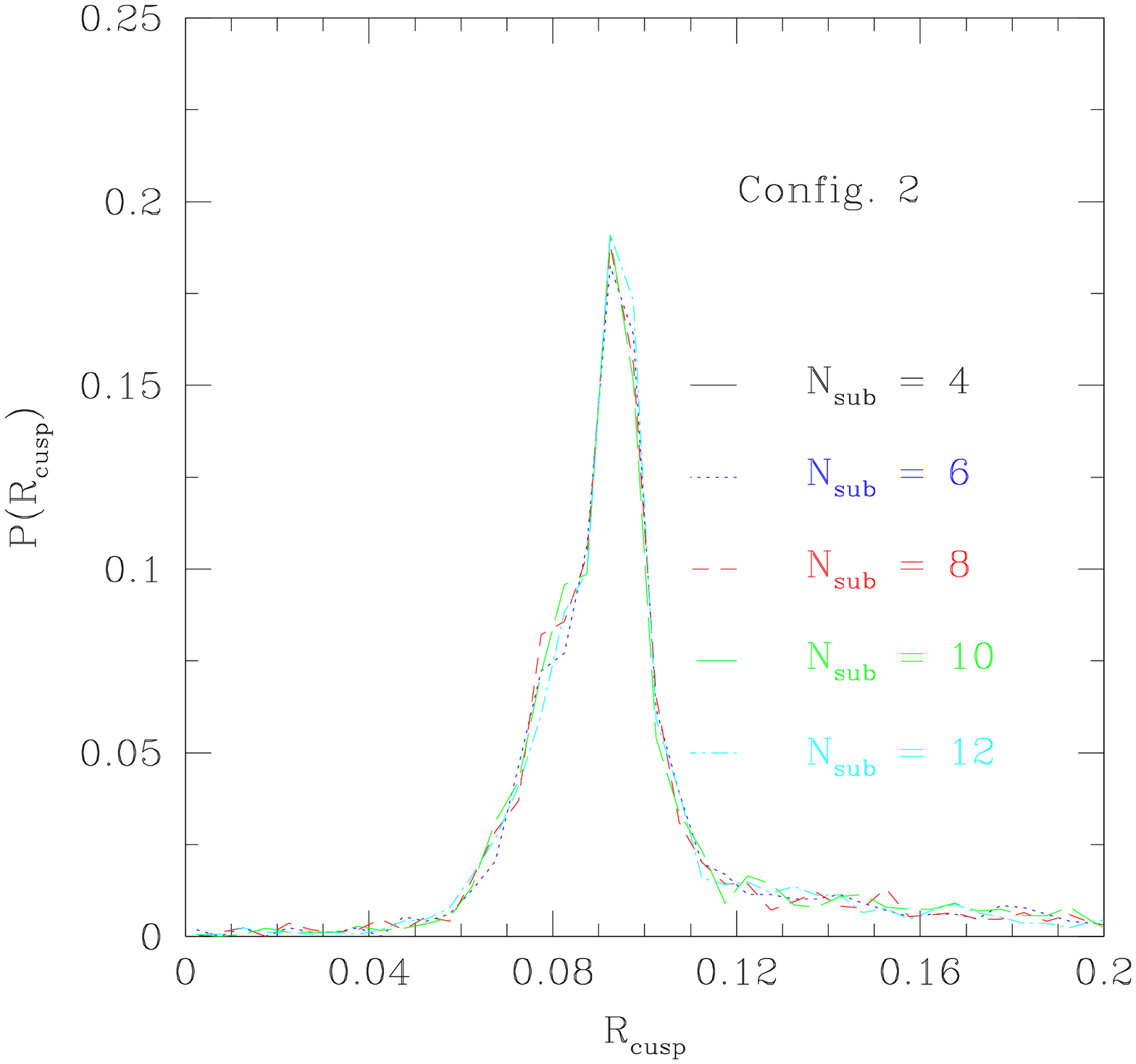}
\caption{Probability distribution of R-variation for config 2, for a different
  number of substructures inside $A_6$.}
\label{fig:Multisub2}
\end{figure}

\begin{figure}
\centering
\plotone{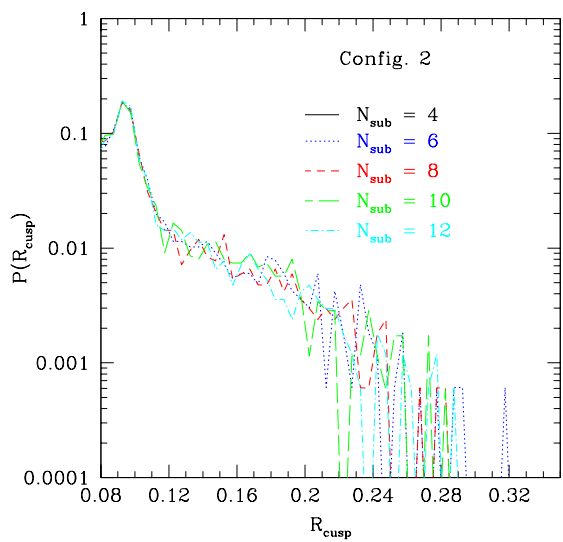}
\caption{Same of figure \ref{fig:Multisub2} with logarithmic scale for the y axis.}
\label{fig:Multisub2big}
\end{figure}

Figure \ref{fig:Multisub2} shows the analogous of figure \ref{fig:Multisub1} for
Config2 and figure \ref{fig:Multisub2big} shows the tail of the distribution for large
values of $\Rc$. 
Even in this case the maximum of the probability distribution is centered
on the unperturbed value. Here the distribution of the values of $\Rc$ is more
symmetric than for Config1 and it is more evident that the effect of substructures
not only increase the value of the cusp ratio but can also reduce it.

%**********************************************************************
%**********************************************************************
%**********************************************************************
%**********************************************************************
%**********************************************************************

\begin{figure}
\centering
\plotone{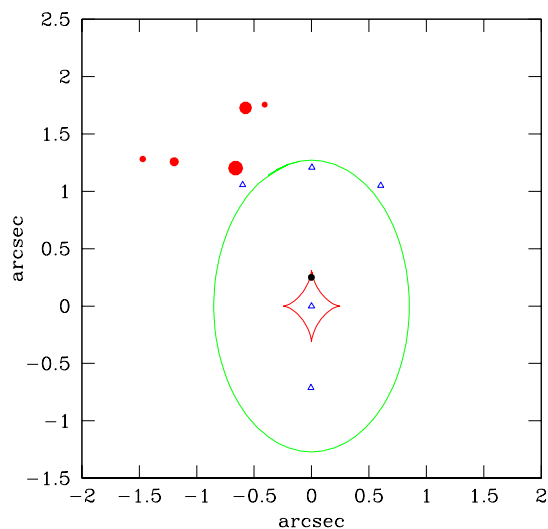}
\caption{One perturbed configuration where $\Rc$ is less than the
  unperturbed value (0.091 $vs$ 0.07). The solid circles show the position of the
  subhaloes, the point size is proportional to their mass.}
\label{fig:lowrc}
\end{figure}

To better illustrate this effect we have isolated one of the configurations
in which we find a reduction of $\Rc$ (figure \ref{fig:lowrc}).
When the distribution of subhaloes is non symmetric in respect to the triplets
of images and one of the perturbers is close to one of the external images
the latter image results to be more magnified than the others. 
In the unperturbed configuration
$|\mu(B)| > \mu(A)+\mu(C)$ and this causes $\Rc\ne 0$. On the other hand if
the perturbers increase $\mu(A)$ without changing considerably the magnification
of the other images, this will enhance the sum $\mu(A)+\mu(C)$ pushing it
closer to $\mu(B)$, giving a smaller $\Rc$ (0.07 in the case of figure \ref{fig:lowrc}).

%**********************************************************************
%**********************************************************************

\begin{figure}
\centering
\plotone{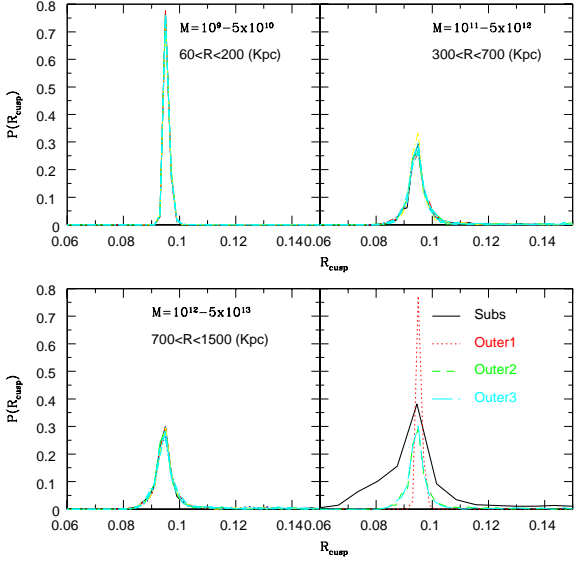}
\caption{Probability distribution of R-variation considering Extra Haloes
  with different masses and distances from the primary lens (Config2). 
  Bottom right panel: comparison between the effects of extra haloes and subhaloes.}
\label{fig:extra2}
\end{figure}

Figure \ref{fig:extra2} shows the effects of extra haloes on the $\Rc$ value
for Config2. For all the three extra-haloes mass ranges considered, the
modifications in the cusp relation are very small and the value of $\Rc$ 
is not very sensitive to the total number of extra-haloes we generated 
(results for different number of subhaloes are shown by different 
curves, which are almost overlapping in the various panels).

%**********************************************************************
%**********************************************************************

\begin{figure}
\plotone{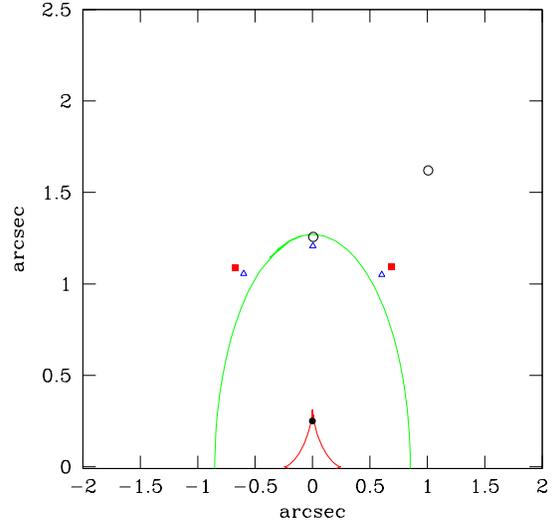}
\caption{Two subhaloes configurations with high $\Rc$ value. The first one is indicated
  by solid squares and gives $\Rc = 0.387$. The second (one indicated by open
  circles) gives $\Rc = 0.274$. In both cases $M_{sub}\approx 10^6$. }
\label{fig:test2sub}
\end{figure}

For this configuration we have performed one more test: we have modified by hand the
fluxes of the three images in order to obtain an high value of $\Rc$ ($>0.37$).
Then we used the {\it gravlens} software to find positions and masses of two
subhaloes (with masses $ 5 \times 10^5 < M < 5 \times 10^6 \Msun$) 
with the constrain of simultaneously reproducing positions and fluxes of
our modified images. We have found two configurations for subhaloes that are shown 
in figure \ref{fig:test2sub}. In the first case (solid squares) the mass of
the substructures is roughly the same ($\approx 2.0 \times 10^6$), they are
close to the external images with a distance of 0.08 arcsec
and the perturbed value of $\Rc$ is 0.387. In the
second case (open circles) the mass of the subhalo close to the central
image is $1.6 \times 10^5 \Msun$ with a distance of 0.06 arcsec, 
while $7.8 \times 10^5 \Msun$ is the mass of the second one that is far
away from the central image, for a perturbed value of $\Rc$ equal to 0.372.

The aim of this test is to show that there is nearly always the possibility to explain
an anomalous flux ratio using subhaloes, but that their positions and masses 
must be tuned in a very precise way (i.e distances between images and subhaloes
must be less than 0.08 arcsec). 
Most important, our Monte Carlo simulations show that the probability of 
obtaining such a fine tuning is very low.
As a consequence we conclude that an explanation for the high number 
of observed anomalous flux ratios in lensed QSOs based on the presence
of subhaloes in the mass range we have tested is very unlikely.

%%%%%%%%%%%%%%%%%%%%%%%%%%%%%%%%%%%%%%%%%%%%%%%%%%%%%%%%%%%%%%%%%%%%%%%%%%%%%%%%55
%%%%%%%%%%%%%%%%%%%%%%%%%%%%%%%%%%%%%%%%%%%%%%%%%%%%%%%%%%%%%%%%%%%%%%%%%%%%%%%%%%%%%

\subsection{Configuration 3}

\begin{figure}
\centering
\plotone{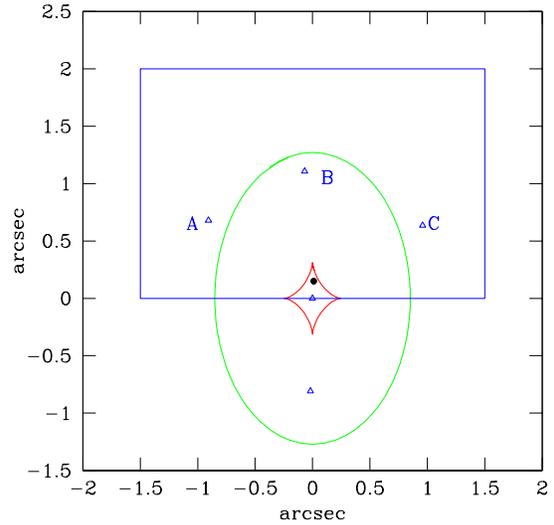}
\caption{Third unperturbed lens configuration: Config3 (R=0.243)}
\label{fig:conf3}
\end{figure}

\begin{figure}
\centering
\plotone{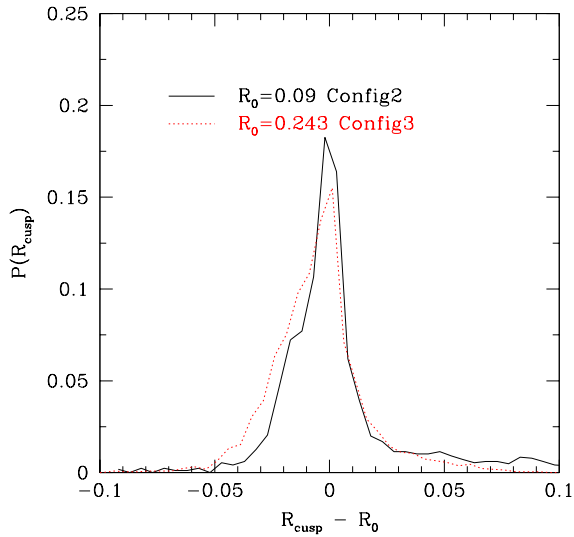}
\caption{Probability distribution for 4 subhaloes within $A_6$ for Config2 and Config3}
\label{fig:Rconfall}
\end{figure}

In the last cusp configuration analyzed (Config3) the R-cusp relation is not
completely satisfied even in the unperturbed case ($\Rc = 0.243$, fig \ref{fig:conf3}).
Figure \ref{fig:Rconfall} shows the R-cusp probability distribution for Config2
and Config3 normalized to the unperturbed value. 
There are no appreciable differences between the two configurations. 
The probability distributions have almost the same width and the
same maximum value ($\approx 0.35$) and both are centered on the respective
unperturbed value. This means that the ability of substructures to modify 
the $\Rc$ relation is nearly independent on the original value of the cusp ratio.

%**********************************************************************
%**********************************************************************
%**********************************************************************

\section{Fold Relation}
\label{sec:fold}

\begin{figure}
\plotone{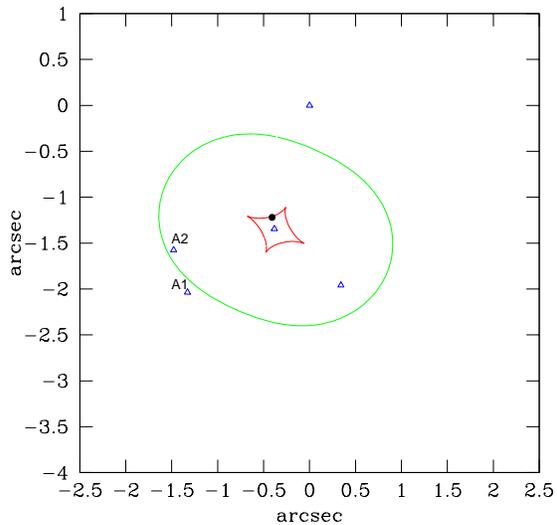}
\caption{Unperturbed fold lens configuration (similar to PG1115, $R_{fold}=0.92$)}
\label{fig:fold1}
\end{figure}

\begin{figure}
\plotone{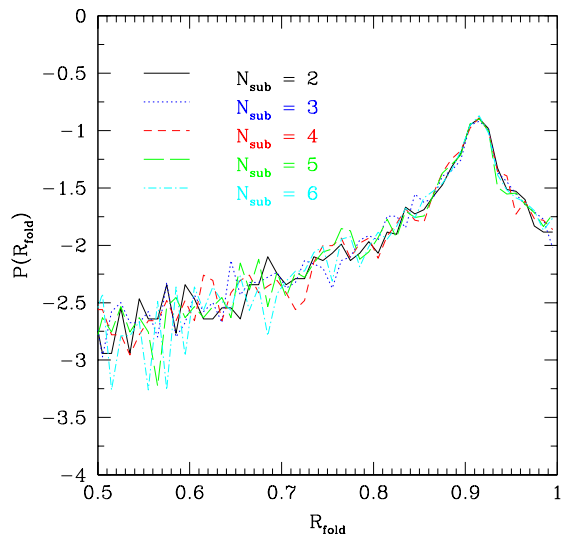}
\caption{Probability distribution of $R_{fold}$ for a different number of
  substructures within an area of 3 arcsec$^2$ ($5\times 10^5 < M_{sub} < 10^7 \Msun$).}
\label{fig:fold2}
\end{figure}

For sake of completeness we have also considered a fold case similar to the
configuration of the well known system PG1115 (Impey et al. 1998).
In this case we define the fold relation as the ratio between the
magnification of the closest pair on opposite sides of the critical curve (cfr
fig \ref{fig:fold1}):
\be
R_{fold} = { \mu(A_1) \over \mu (A_2) }=1.
\ee
By applying our procedure to this test fold case we found that it is easier to
modify the value of the unperturbed $R_{fold}$ with respect to the one of $\Rc$,
as clearly shown in fig \ref{fig:fold2}.
%AM
In our computational procedure the position of the source can be changed by
the {\it gravlens} code, in order to minimize the $\chi^2$ of the lens
configuration. While this does not affect the $\Rc$ relation it can be
important in the fold case:
%AM
as pointed out by Keeton, Gaudi and Petters (2005) the degree to which
$R_{fold}$ can differ from one for realistic smooth lenses depends, in
addition to the angular structure of the lens potential, not only on 
the distance of the source from the caustic but also on its location along
the caustic itself.
%AM
So the the values we got for $R_{fold}$ are due both to the effects of
subhaloes and both to the shift along the caustics of the source.

Thus, it is not possible to conclude from $R_{\rm fold}$ alone whether observed flux ratios
are anomalous or not.

%%%%%%%%%%%%%%%%%%%%%%%%%%%%%%%%%%%%%%%%%%%%%%%%%%%%%%%%%%%%%%%%%%%%%%%%%%%%%%%%55
%%%%%%%%%%%%%%%%%%%%%%%%%%%%%%%%%%%%%%%%%%%%%%%%%%%%%%%%%%%%%%%%%%%%%%%%%%%%%%%%55

%%%%%%%%%%%%%%%%%%%%%%%%%%%%%%%%%%%%%%%%%%%%%%%%%%%%%%%%%%%%%%%%%%%%%%%%%%%%%%%%55
%%%%%%%%%%%%%%%%%%%%%%%%%%%%%%%%%%%%%%%%%%%%%%%%%%%%%%%%%%%%%%%%%%%%%%%%%%%%%%%%%%%%%

\section {Discussion \& Conclusions}
\label{sec:concl}

Quasars that are being gravitationally lensed into multiple images have
recently been used to place limits on the surface density of cold dark matter subhaloes
(Mao \& Schneider 1998, Metcalf \& Madau 2001, Chiba 2002,
Metcalf \& Zhao 2002, Dalal \& Kochanek 2002, Chen Kravtsov \& Keeton 2003,
Bradac et al. 2004, Mao et al. 2004). 
Small mass clumps that happen to lie near the images affect the observed magnification
ratios. The question arises as to whether these observations are
compatible with distortions expected to occur from dark matter substructures
and satellite galaxies within the $\Lambda$CDM model.
Recent results based on numerical N-body (Amara et al. 2004, Rozo et al. 2005) 
and hydro simulations (Macci\`o et al. 2005) have shown that it is 
hard to reconcile the observed
high number of cups relation violation with the total amount of substructures
predicted by the \LCDM model.
These studies were limited by the present achievable numerical resolution
that permits to resolve DM haloes down to masses $\approx 10^7 \Msun$.

In this work we have quantified the effects of smaller mass clumps
($10^5-10^7 \Msun$) on the observed violation of the R cusp relation.
We employed results from N-body simulations to estimate the expected number
of subhaloes in this low mass range. Due to the small mass of the perturbers we
have restricted our analysis only to those close (in 2D) to the
images positions. For the mass range inspected in this work and for the
typical distance between images (few arcsecs) 
this leads to a number of perturbers $\approx 6$. All the subhaloes are
modeled as NFW spheres and we have generated more than $10^5$ different
lensing configurations, varying masses, positions and number of subhaloes.

The main finding of our work is that on a statistical basis
this class of perturbers is not able to modify consistently the unperturbed R-cusp relation. 
Values of $\Rc$ in the observed range ($\approx 0.25$) are obtained in only
less than 1\% of the analyzed systems.  
The ability of subhaloes in modifying the unperturbed value of
$\Rc$ is found to be independent from the value of $\Rc$ itself.

These results are not in contradiction with the ones in the literature (Keeton
et al. 2003, and more recently Miranda \& Jetzer 2005 and references therein).
As shown in figure \ref{fig:test2sub} it is possible to use
subhaloes in the mass range $10^5-10^7 \Msun$ to obtain high values of $\Rc$
case by case, but a tight {\it fine tuning} between the location of
the images and masses/positions of the perturbers is needed . 

In addition, we have also considered the impact of massive haloes placed outside 
the primary lens (from groups of galaxies to a close cluster) by modeling them
in the same way of the subhaloes. 
Our simulations show, as expected, that their contribution in
modifying the R cusp relation is tiny and almost negligible with respect to the
effect of subhaloes.

Results from this work together with results from numerical
simulations seem to be in disagreement with the standard picture which explains
the anomalous flux ratio by means of dark matter satellites. 
Interestingly, while on dwarf galaxy scale there is an excess of dark matter subhaloes
with respect to visible satellites, we have shown that the predicted level of
substructures on smaller scales is not sufficient to explain the observed level
of violation in the cusp relation.

Possible solutions to this problem can reside in microlensing for some of the
lensing systems observed in the optical band (Metcalf 2005, Keeton et
al 2005), or in the presence of haloes lying along the line of sight between 
the lens and the observer (Chen et al. 2003, Metcalf 2005), 
although total effect of this kind of perturbers is not yet clear. 

\section*{Acknowledgments}

We thank Chuck Keeton for making available the {\it gravlens} software 
and for his comments on an early version of this work.
We also thank J. Diemand for making available his Nbody simulations
and M. Bartelmann, Ph. Jetzer and R. Piffaretti for useful discussions.
An anonymous referee is also thanked for his suggestions that improved the
presentation of this work.
Marco Miranda was partially supported by the Swiss National Science Foundation.

\end{document}